\documentclass[11pt,twoside,a4paper,twocolumn]{article}
\usepackage{graphicx}
\usepackage{appendix}
\usepackage{txfonts}

\begin{document}
\title{Investigation of HNCO isomers formation in ice mantles by UV and thermal processing: an experimental approach}
\author{A. Jim\'enez-Escobar $^1$, B. M. Giuliano $^1$, G. M. Mu\~{n}oz Caro $^1$, J. Cernicharo $^1$, N. Marcelino $^2$\\
 $^1$Centro de Astrobiolog\'{\i}a, INTA-CSIC, Carretera de Ajalvir, km 4, Torrej\'on de Ardoz, 28850 Madrid, Spain\\
 $^2$National Radio Astronomy Observatory, 520 Edgemont Road, Charlottesville, VA 22903, USA\\
\texttt{bgiuliano@cab.inta-csic.es}}
\date{June 2014}
\maketitle

%\title{Investigation of HNCO isomers formation in ice mantles by UV and thermal processing: an experimental approach}
%\author{A. Jim\'enez-Escobar \altaffilmark{1}, B. M. Giuliano \altaffilmark{1}, G. M. Mu\~{n}oz Caro \altaffilmark{1}, J. Cernicharo \altaffilmark{1}, N. Marcelino \altaffilmark{2}}
%\altaffiltext{1}{Centro de Astrobiolog\'{\i}a, INTA-CSIC, Carretera de Ajalvir, km 4, Torrej\'on de Ardoz, 28850 Madrid, Spain}
%\altaffiltext{2}{National Radio Astronomy Observatory, 520 Edgemont Road, Charlottesville, VA 22903, USA}

%\email{bgiuliano@cab.inta-csic.es}

%\received{2013}
%\revised{2013}
%\accepted{2013}

\begin{abstract}

 Current gas phase models do not account for the abundances of HNCO isomers detected in various environments, suggesting a formation in icy grain mantles. We attempted to study a formation channel of HNCO and its possible isomers by vacuum-UV photoprocessing of interstellar ice analogues containing H$_2$O, NH$_3$, CO, HCN, CH$_3$OH, CH$_4$, and N$_2$ followed by warm-up, under astrophysically relevant conditions. Only the H$_2$O:NH$_3$:CO and H$_2$O:HCN ice mixtures led to the production of HNCO species. The possible isomerization of HNCO to its higher energy tautomers following irradiation or due to ice warm-up has been scrutinized. The photochemistry and thermal chemistry of H$_2$O:NH$_3$:CO and H$_2$O:HCN ices was simulated using the Interstellar Astrochemistry Chamber (ISAC), a state-of-the-art ultra-high-vacuum setup. The ice was monitored in situ by Fourier transform mid-infrared spectroscopy in transmittance. A quadrupole mass spectrometer (QMS) detected the desorption of the molecules in the gas phase. UV-photoprocessing of H$_2$O:NH$_3$:CO/H$_2$O:HCN ices lead to the formation of OCN$^-$ as main product in the solid state and a minor amount of HNCO. The second isomer HOCN has been tentatively identified. Despite its low efficiency, the formation of HNCO and the HOCN isomers by UV-photoprocessing of realistic simulated ice mantles, might explain the observed abundances of these species in PDRs, hot cores, and dark clouds.

\end{abstract} 


%%%%%%%%%%%%%%%%%%%%%%%%%%%%%%%%%%%%%%%%%%%%%%%%%%%%%%%%%%
\section{Introduction}
\label{intro}
\begin{figure} [ht!]
   \centering
    \includegraphics[angle=0,width=4.5 cm]{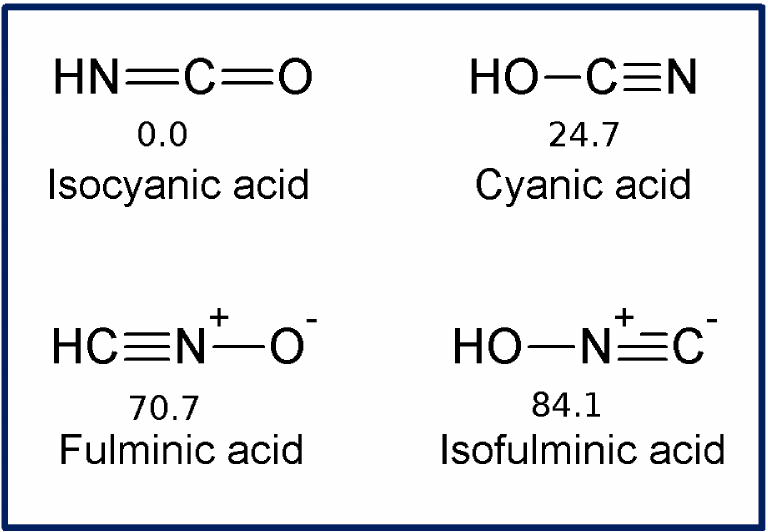}   
      \caption{Sketch of the HNCO isomers and their relative energies expressed in kcal/mol (taken from Br\"{u}nken et al. 2009)}
      \label{isomers}
\end{figure}

\indent\indent The detection of isocyanic acid (HNCO) and its isomers in different astrophysical environments is difficult to explain with current gas phase models, which suggests that these species could be formed in ice mantles and desorb to the gas phase (Quan et al. 2010, Marcelino et al. 2010). HNCO has been observed in a wide range of physical conditions, from diffuse and cold dark clouds to dense cores in star forming regions, but also toward the warm molecular clouds in the Galactic Center and external galaxies (see Buhl et al. 1973 and Nguyen-Q-Rieu et al. 1991). This species has several metastable isomers, namely HOCN, HCNO, and HONC, lying at increasingly higher energy than HNCO (see Mladenovi\'c \& Lewerenz 2008, and Mladenovi\'c et al. 2009). In Fig.~\ref{isomers} is shown the sketch of the four stable HNCO isomers. While HOCN and HCNO have been both observed toward dark clouds, in warmer regions, such as Sgr B2 and the low-mass protostar IRAS 16293-2422, only HOCN has been detected (see Marcelino et al. 2009, 2010 and ref. therein; Br\"unken et al. 2010). The gas-phase modeling in Marcelino et al. (2010) could reproduce reasonably well the results on cold clouds, but, for warmer environments, a gas-grain model with a warm-up phase was introduced. The production of HNCO and HOCN on dust grain surfaces has also been explored by Br\"unken et al. (2010) in order to explain the observed abundances toward Sgr B2. All these results motivated a new gas-grain modeling of the CHNO isomers by Quan et al. (2010). These studies concluded that a combination of surface and gas-phase chemistry is needed for the production of the CHNO isomers, as was previously found by Tideswell et al. (2010) for the lowest HNCO isomer. It is, however, not clear how the HNCO isomers could be formed in ice mantles. New laboratory simulations are thus required to study their formation pathways in the solid phase.

Grim et al. (1989), Lowenthal et al. (2000), and Raunier et al. (2003) have observed the thermal deprotonation of HNCO in ice analogues by, respectively, ammonia and water, and the consequent formation of the cyanate ion (OCN$^-$).
An infrared absorption at 4.62 $\mu$m (2165 cm$^{-1}$) observed toward several protostars, is attributed to OCN$^-$ in the ice (Grim et al. 1989, Schutte \& Greenberg 1997, and Novozamsky et al. 2001 and ref. therein). Its non-detection in dark clouds (Knez et al. 2005) suggests that this species is the result of irradiation or thermal processing.
In ice processing experiments, OCN$^-$ forms in various ways: proton bombardment (Moore et al. 1983), electron bombardment (Bennett et al. 2010), vacuum-UV (VUV) photolysis of some ice mixtures (Grim et al. 1989), or by acid-base reactions including HNCO among the reactants (Raunier et al. 2003, and van Broekhuizen et al. 2004). The kinetics for the HNCO + NH$_3$ reaction has been reported recently by Mispelaer et al. (2012), while the kinetics for OCN$^-$ and the formation of HOCN isomer by thermal processing, based on a very weak infrared absorption band, was reported by Theule et al. (2011). A first reaction deprotonates HNCO into OCN$^-$, and a second reaction protonates OCN$^-$ into HNCO and HOCN. The ice surface helps transferring the proton from the nitrogen atom to the oxygen atom (Theule et al. 2011).

In UV-irradiation experiments the formation of HNCO is followed by a proton transfer to NH$_3$ (Grim et al. 1989). We attempted to study the formation of OCN$^-$ and HNCO by photoprocessing of ice mixtures containing H$_2$O, NH$_3$, CO and HCN, followed by warm-up, under astrophysically relevant conditions in a ultra-high vacuum set-up. We also attempted to investigate the possible formation of the high-energy isomers HOCN, HCNO, and HONC under similar conditions in the ice. For that, we used the ISAC set-up at Centro de Astrobiolog\'ia (CAB) in Madrid, using infrared and quadrupole mass spectroscopy to monitor the solid and gas phases, respectively.
%%%%%%%%%%%%%%%%%%%%%%%%%%%%%%%%%%%%%%%%%%%%%%%%%%%%%%%%%%
\section{Experimental}\label{expe}

\indent\indent Ice analogues have been prepared from different gas mixtures by condensation onto a KBr window cooled at 7 $\mathrm{K}$, placed in the Interstellar Astrochemistry Chamber (ISAC) set-up described in  Mu\~{n}oz Caro et al. (2010). ISAC is an ultra high vacuum (UHV) set-up, with base pressure down to P=2.5-4.0 $\times$ 10$^{-11}$ mbar measured at room temperature, after baking the set-up. The deposited ices have been processed by VUV irradiation using a microwave-stimulated hydrogen flow discharge lamp that provides a flux of 2.5 $\times$ 10$^{14}$ photons cm$^{-2}$ s$^{-1}$. The lamp emits in the 7.3-10.5 eV (169.8-118.1 nm) range, an average photon energy of 8.6 eV (144.2 nm), with  maxima of emission  at Lyman-$\alpha$ around 10.20 eV (121.6 nm), and the Lyman band system of molecular emission with peaks at 157.8 nm (7.85 eV) and 160.8 nm (7.71 eV), see Mu\~{n}oz Caro et al. (2010), Cruz-Diaz (2014), and Chen et al. (2014).

The irradiated ice samples have been subsequently heated to room temperature in a controlled way, allowing Temperature Programmed Desorption (TPD) experiments. The evolution of the ice samples has been monitored by Fourier transform infrared (FTIR) transmittance spectroscopy, while the desorption products have been analyzed by mass spectrometry (MS) using a quadrupole mass analyzer (Pfeiffer Vacuum Prisma QMS 200 with a Channeltron detector) with an electron impact energy of 70 eV.

Gaseous mixtures of H$_2$O (triply distilled), CO, NH$_3$ (both supplied by Praxair, 99.998\% and 99.999\% pure, respectively) and HCN have been prepared in a gas line in which the entrance of the individual components is controlled by commercial Pfeiffer Vacuum electrical valves. The partial pressures of the gases was measured by a quadrupole mass spectrometer (Pfeiffer Vacuum Prisma QMS 200 with a Faraday detector) connected to the line.

HCN has been synthesized from potassium cyanide and stearic acid using the same method described in Gerakines et al. (2004). The solid mixture has been heated up to 330 $\mathrm{K}$ and the gaseous reaction products have been collected in a glass vessel connected to the gas line.

The column density of the deposited ice is calculated using the formula\\
\begin{equation}
 N =\frac{1}{A}\int_{band}{\tau_{\nu}d\nu}
\label{column}
\end{equation}
where $N$ is the column density in cm$^{-2}$, $\tau_{\nu}$ the optical depth of the band, $d\nu$ the wavenumber differential in cm$^{-1}$, and $A$ the band strength in cm molecule$^{-1}$. The integrated absorbance is equal to 0.43 $\tau$, where $\tau$ is the integrated optical depth of the band.

The adopted band strengths for HCN, NH$_3$, CO and H$_2$O are $A$(HCN) = 5.1~ $\times$ ~10$^{-18}$ cm molecule$^{-1}$ at 4.76 $\mu$m (2100 cm$^{-1}$) from Bernstein et al. (1997), $A$(NH$_3$) = 1.7 $\times$  10$^{-17}$ cm molecule$^{-1}$ at 9.35 $\mu$m (1070 cm$^{-1}$) from Sandford \& Allamandola (1993), $A$(CO) = 1.1 $\times$  10$^{-17}$ cm molecule$^{-1}$ at 4.68 $\mu$m (2139 cm$^{-1}$) from Jiang et al. (1975), and $A$(H$_2$O) = 2.0 $\times$ 10$^{-16}$ cm molecule$^{-1}$ at 3.05 $\mu$m (3279 cm$^{-1}$) from Hagen et al. (1981). Typical uncertainties  in the estimated band strengths can vary from 5\% up to 50\% depending on the ice molecular composition and analysis, as reported by Gerakines et al. (1995).
    
The UV photon fluence, in photon cm$^{-2}$, is the product of the UV flux by the irradiation time. The average UV dose experienced by the ice layer, in photon molecule$^{-1}$, was calculated dividing the UV photon fluence by the ice column density.

In the H$_2$O:HCN mixture experiments the deposition and irradiation processes were repeated four times with a similar ice thickness in order to increase the concentration of the HNCO species. Each layer was deposited on top of the previous one and irradiated. This experimental procedure is necessary to ensure the homogeneous irradiation of thicker ices, due to the small penetration depth of the UV photons.

\begin{figure} [ht!]
   \centering
    \includegraphics[width=7.5 cm]{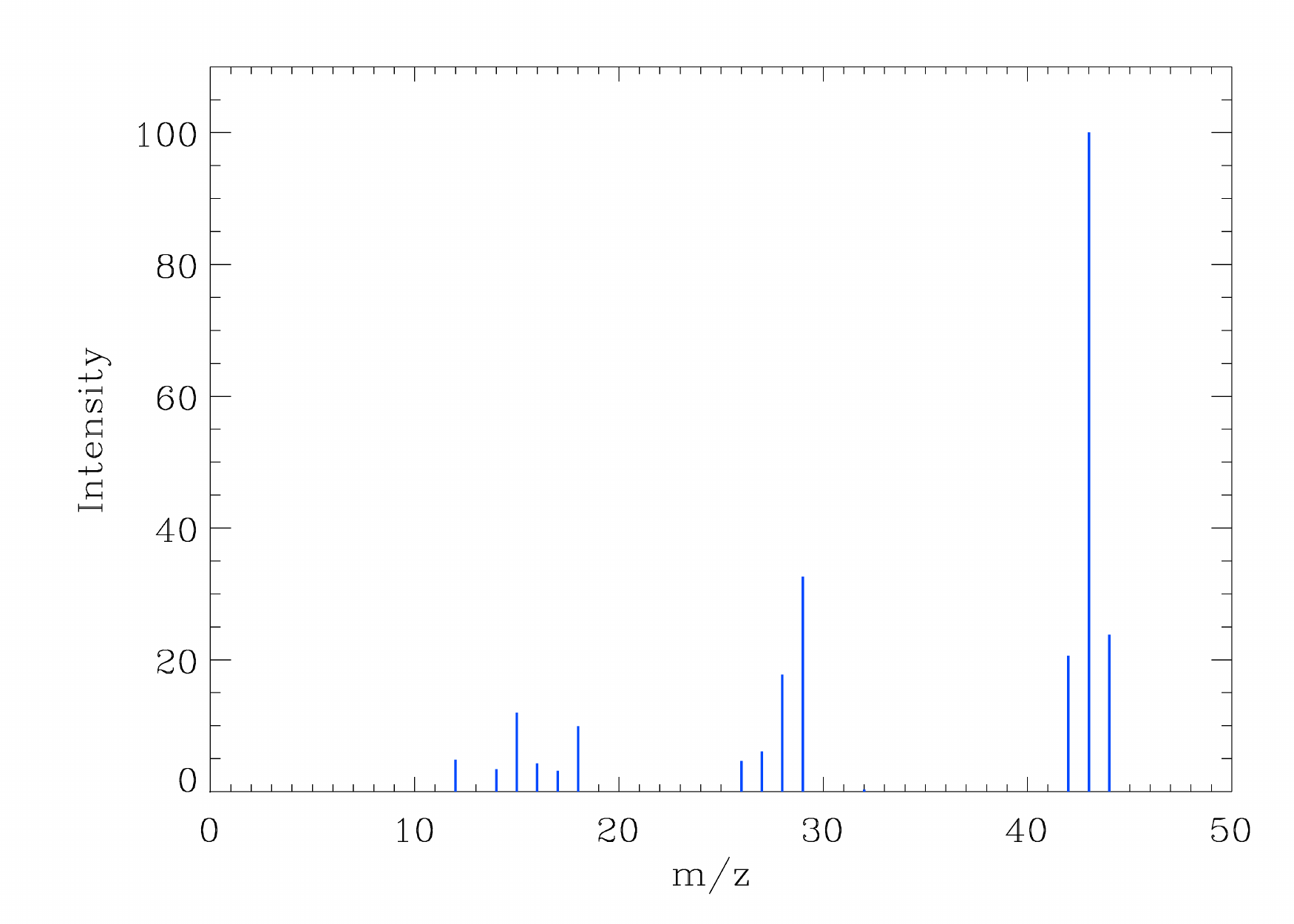}   
      \caption{Mass spectrum of pure gas phase HNCO recorded with an electron impact (70 eV) QMS.}
      \label{fragm}
\end{figure}

Fischer et al. (2002) report a mass spectrometric study about the fragmentation pattern of the HNCO species in the electron ionization mode. This study assigns the most abundant observed fragment to the NCO$^+$ species, with a relative intensity to the molecular ion of 20\%. We performed similar experiments on the pure HNCO (solid polymer, heated up to about 700 $^o$C to vaporize it). The mass spectrum has been recorded and compared with the one presented in Fischer et al. (2002), indicating that the fragmentation ratio observed in our experiment is consistent with the one presented by Fischer et al. (2002). The recorded spectrum is shown in Fig.~\ref{fragm}.
In the spectrum some lines due to the contamination of the polymer are also visible. It was not possible to generate and analyze the HOCN isomer in our experimental set-up due to its high instability; its mass spectrum was not found in the literature.

%%%%%%%%%%%%%%%%%%%%%%%%%%%%%%%%%%%%%%%%%%%%%%%%%%%%%%%%%%%%%
\section{Experimental results}
\label{results} 
\indent\indent The experimental results of deposition, irradiation and subsequent heating are described below.  Sect.~\ref{results_1} shows the results of an ice mixture of astrophysical relevance, made by co-deposition of water, ammonia and carbon monoxide in a 2:1:1 ratio, while Sect.~\ref{results_2} shows the results of an ice mixture containing water and hydrogen cyanide in a 2:1 ratio that optimizes the synthesis of HNCO isomers.

In addition to the experiments presented in the next sections, the formation of the HNCO species and its isomers has also been investigated using different ice mixtures, containing molecules of astrophysical relevance. In particular, we considered ice mixtures of methanol and ammonia; water, methane, and molecular nitrogen; and water, molecular nitrogen, and carbon monoxide. None of them led to the production of the HNCO species and its isomers.

\subsection{H$_2$O:NH$_3$:CO ice}
\label{results_1}
\indent\indent The energetic processing (by UV- and thermally-mediated solid state chemistry) and proton irradiation of an ice mixture composed of H$_2$O, NH$_3$ and CO has already been investigated by van Broekhuizen et al. (2004) and Hudson et al. (2001), respectively. Their results evidenced the formation, in the processed ice, of the cyanate ion (OCN$^-$) as main photoproduct, and, in a lesser amount, of isocyanic acid (HNCO). Nevertheless, the production rate of the HNCO species is very low and its spectroscopic features are hardly visible in thin ice matrices made of few monolayers.

\begin{figure} [ht!]
   \centering
    \includegraphics[width=7.5cm]{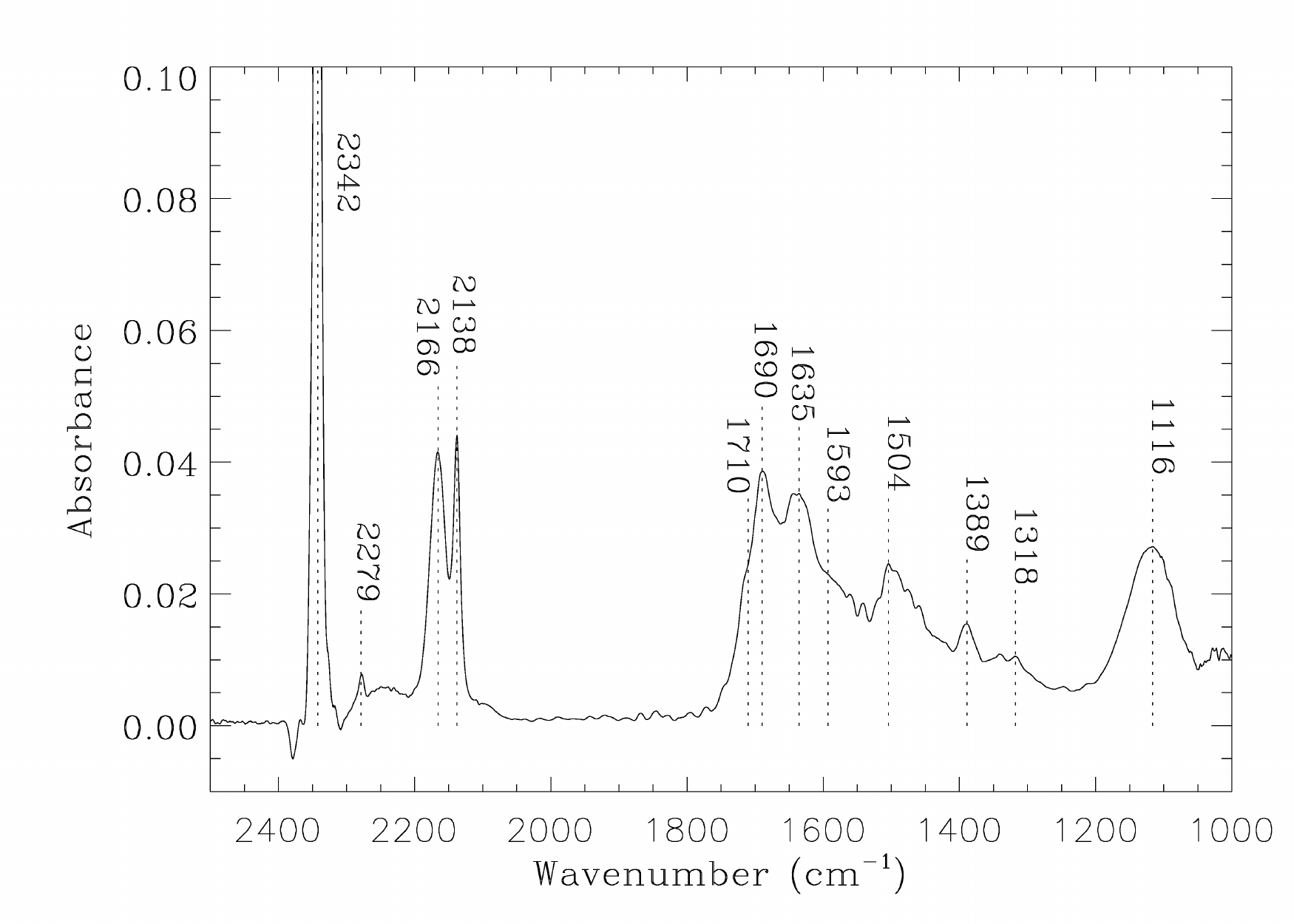} 
     \caption{Infrared spectrum  of H$_2$O:NH$_3$:CO (2:1:1) ice mixture after 90 minutes of simultaneous deposition and irradiation.}
      \label{warmup_ftir}
\end{figure}

The infrared spectrum of a photoprocessed H$_2$O:NH$_3$:CO (2:1:1) ice, deposited at 8 $\mathrm{K}$ after 90 minutes of simultaneous deposition and irradiation, is shown in Fig.~\ref{warmup_ftir}. The estimated dose was 1 photon mol$^{-1}$ on average.
The main HNCO absorption at 2260 cm$^{-1}$, listed in van Broekhuizen et al. (2004), is almost undetectable in our experiment. If HNCO is produced, it will probably react very rapidly to give OCN$^-$ (see Mispelaer et al. 2012). The peak assignment of the ice initial components and photoproducts is listed in Table~\ref{band_position_1}.

\begin{figure} [ht!]
   \centering
    \includegraphics[width=7.5cm]{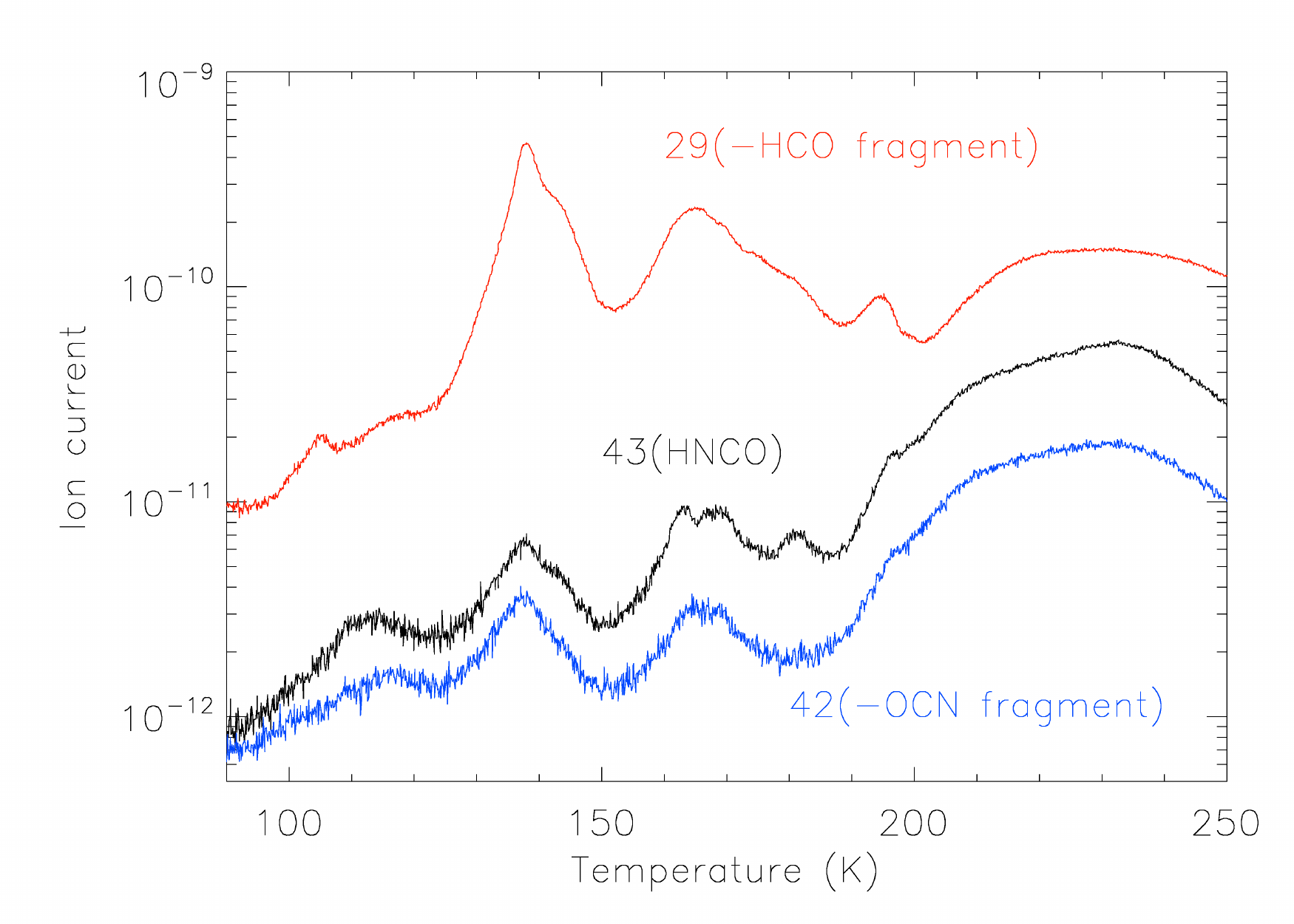}
    \includegraphics[width=7.5cm]{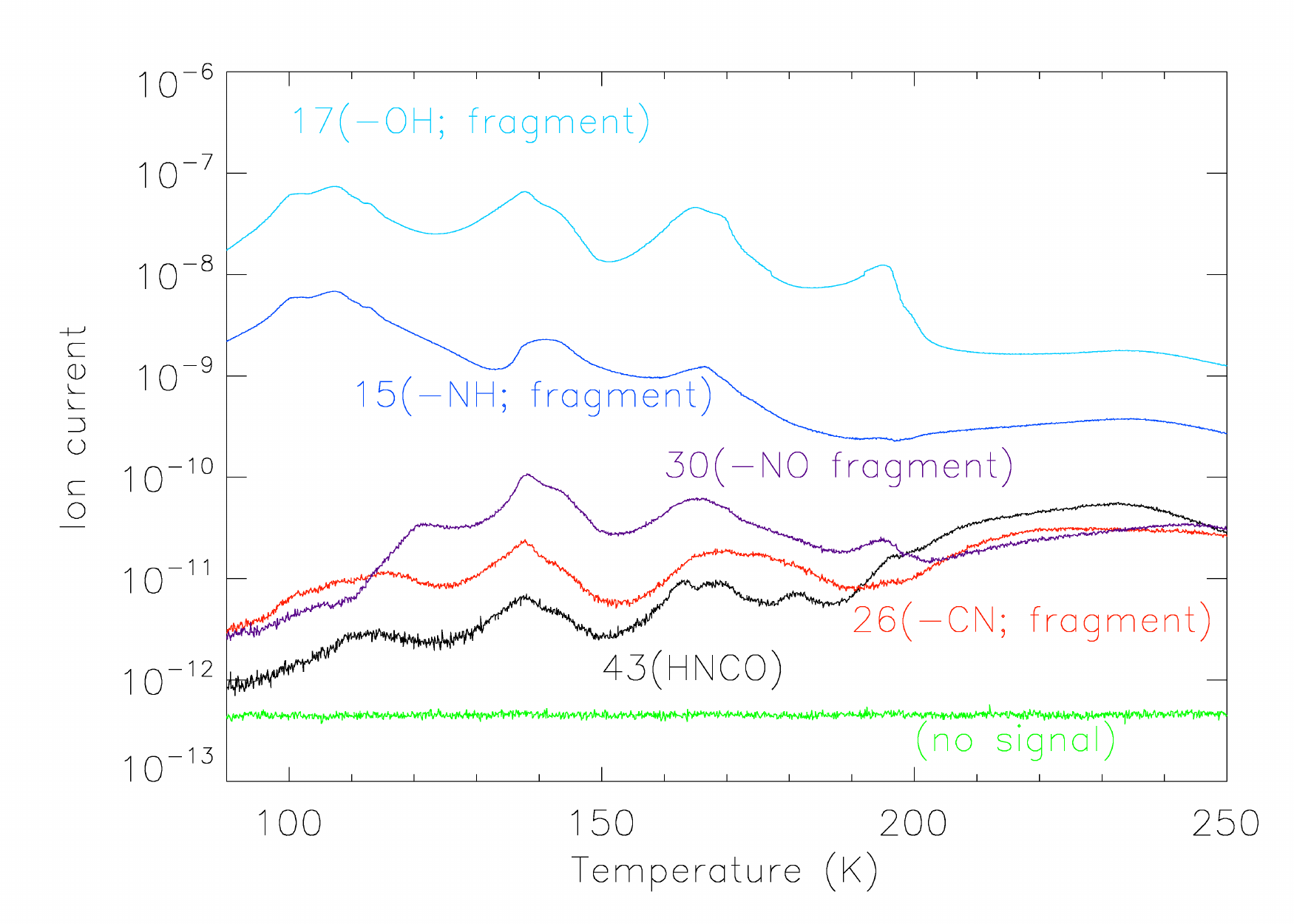}
     \caption{Thermal desorption of UV-irradiated H$_2$O:NH$_3$:CO (2:1:1) ice mixture. The m/z~=~43 value refers to the molecular ion of the HNCO species and its isomers. The figure is split in two panels: the top panel shows the masses corresponding to the most abundant fragments of the HNCO isomer; the bottom panel shows the potential fragments of other isomers that might be present in the ice.}
      \label{warmup_qms}
\end{figure}

However, the presence of HNCO is detectable in the mass spectrum recorded during the thermal desorption of the irradiated ice. Fig.~\ref{warmup_qms} shows the TPD curves, recorded using a heating ramp of 2 $\mathrm{K}$ min$^{-1}$, of the m/z values for the HNCO molecular ion (m/z~=~43) and its fragments NCO$^+$ (m/z~=~42), and HCO$^+$ (m/z~=~29). The peaks observed below 200 $\mathrm{K}$ are due to co-desorption of the photoproducts with NH$_3$ and H$_2$O. The fragmentation pattern can be used to obtain information about the spatial arrangement of the atoms in the molecule. We are mainly interested in the temperature range after water desorption has occured, when the OCN$^-$NH$_4^+$ salt, formed by an acid-base reaction, is expected to desorb as NH$_3$ + HNCO and HOCN, and there is no contribution of the H$_2$O mass fragments in the measurements. The m/z~=~29 intensity appears high, most probably because there is a contribution of other species, like HCONH$_2$ or H$_2$NCONH$_2$ (Raunier et al. 2004), detected in the irradiated ice by IR spectroscopy, see Fig.~\ref{warmup_ftir} and Table~\ref{band_position_1}. Unfortunately, only the mass spectrum of the HNCO isomer is experimentally available, and a reliable prediction of the fragmentation pattern for the other isomers is not straightforward. Nevertheless, we payed attention to m/z~=~17 ($-$OH), and NH$_3$ produced after decomposition and desorption of the OCN$^-$NH$_4^+$ salt, m/z~=~15 ($-$NH), m/z~=~26 ($-$CN), and m/z~=~30 ($-$NO) as potential fragments characteristic of the HNCO and HOCN isomers. A typical curve profile when no ion was detected in the desorbing gas is also shown, indicated by the label (no signal).

%%%%%%%%%%%%%%%%%%%%%%%%%%%%%%%%%%%%%%%%%%%%%
%%%%%%%%%%%%%%%%%%%%%%%%%%%%%%%%%%%%%%%%%%%%%%
%%%%%%%%%%%%%%%%%%%%%%%%%%%%%%%%%%%%%%%%%%%%%%
\begin{table}
\centering
\caption{Infrared band positions in a UV-photoprocessed H$_2$O:NH$_3$:CO (2:1:1) ice in the range of interest from 2500 to 1000 cm$^{-1}$.}
\label{band_position_1}
\begin{tabular}{cc}   \hline \hline
\multicolumn{1}{c}{Wavenumber (cm$^{-1}$)} & Molecule \\ \hline
2342 & CO$_2$\\
2279 & $^{13}$CO$_2$\\
2166 & OCN$^-$\\
2138 & CO\\
1710 & H$_2$CO\\
1690 & HCONH$_2$\\
1635 & H$_2$O\\
1593 & HCOO$^-$\\
1504 & H$_2$CO\\
1485 & NH$_4$$^+$$^a$\\
1389 & HCONH$_2$\\
1318 & HCONH$_2$\\
1116 & NH$_3$\\
\hline
\end{tabular}
\begin{list}{}
\item $^a$ From van Broekhuizen et al. (2004).
\end{list}
\end{table}
%%%%%%%%%%%%%%%%%%%%%%%%%%%%%%%%%%%

%%%%%%%%%%%%%%%%%%%%%%%%%%%%%%%%%%%%%%%%%%%%
\subsection{H$_2$O:HCN ice}
\label{results_2}

%%%%%%%%%%%%%%%%%%%%%%%%%%%%%%%%%%%%%%%%%%%%%%%%%%%%%%%%%%%%%%%%%
\indent\indent The formation of OCN$^-$ and HNCO in irradiated H$_2$O:HCN ice mixtures has been observed by Gerakines et al. (2004). As already mentioned, we moved to the analysis of this ice mixture in order to produce the HNCO species in a more efficient way, thus allowing the identification of potential isomers.

\begin{figure} [ht!]
   \centering
    \includegraphics[width=7.5cm]{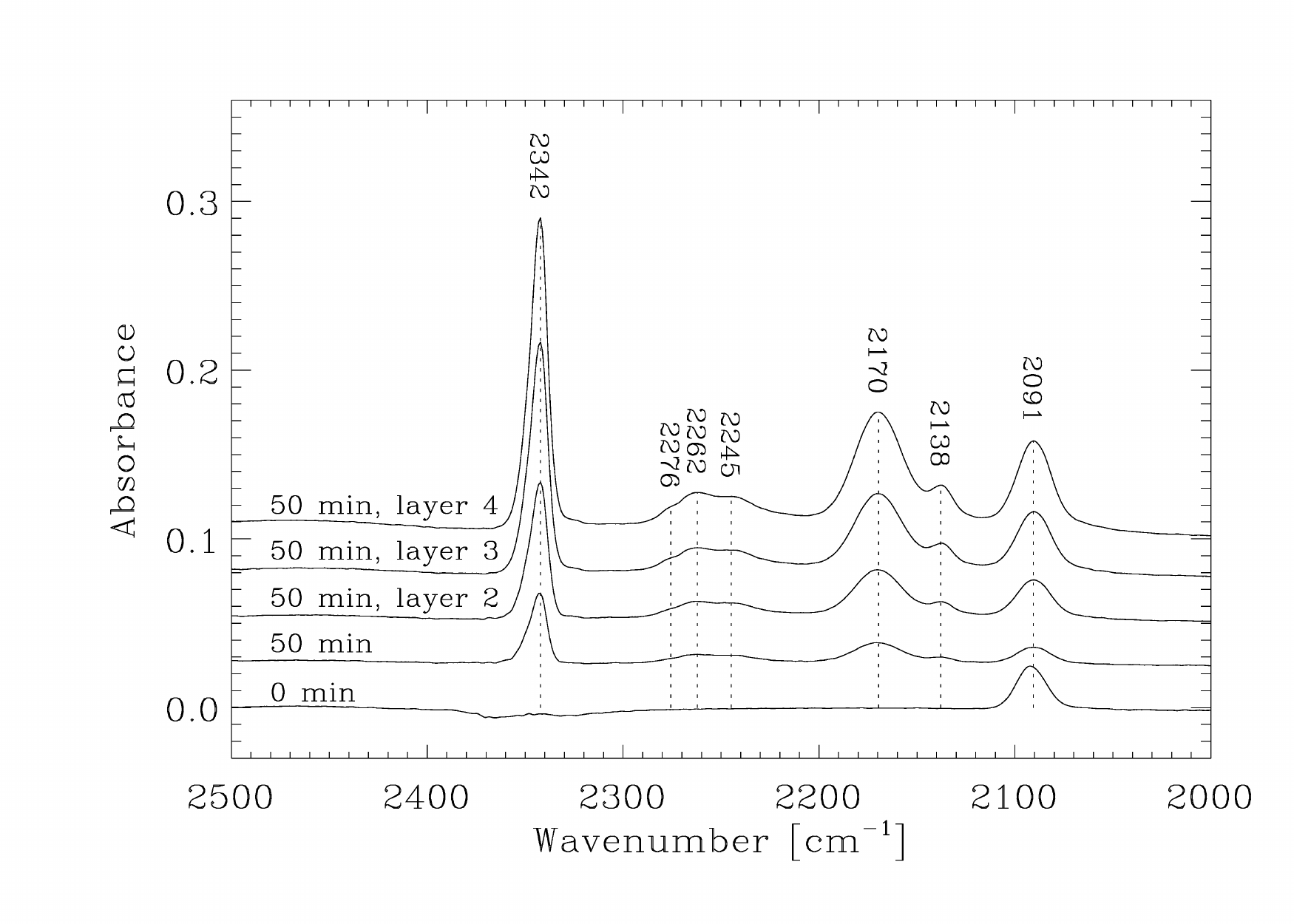} 
    \includegraphics[width=7.5cm]{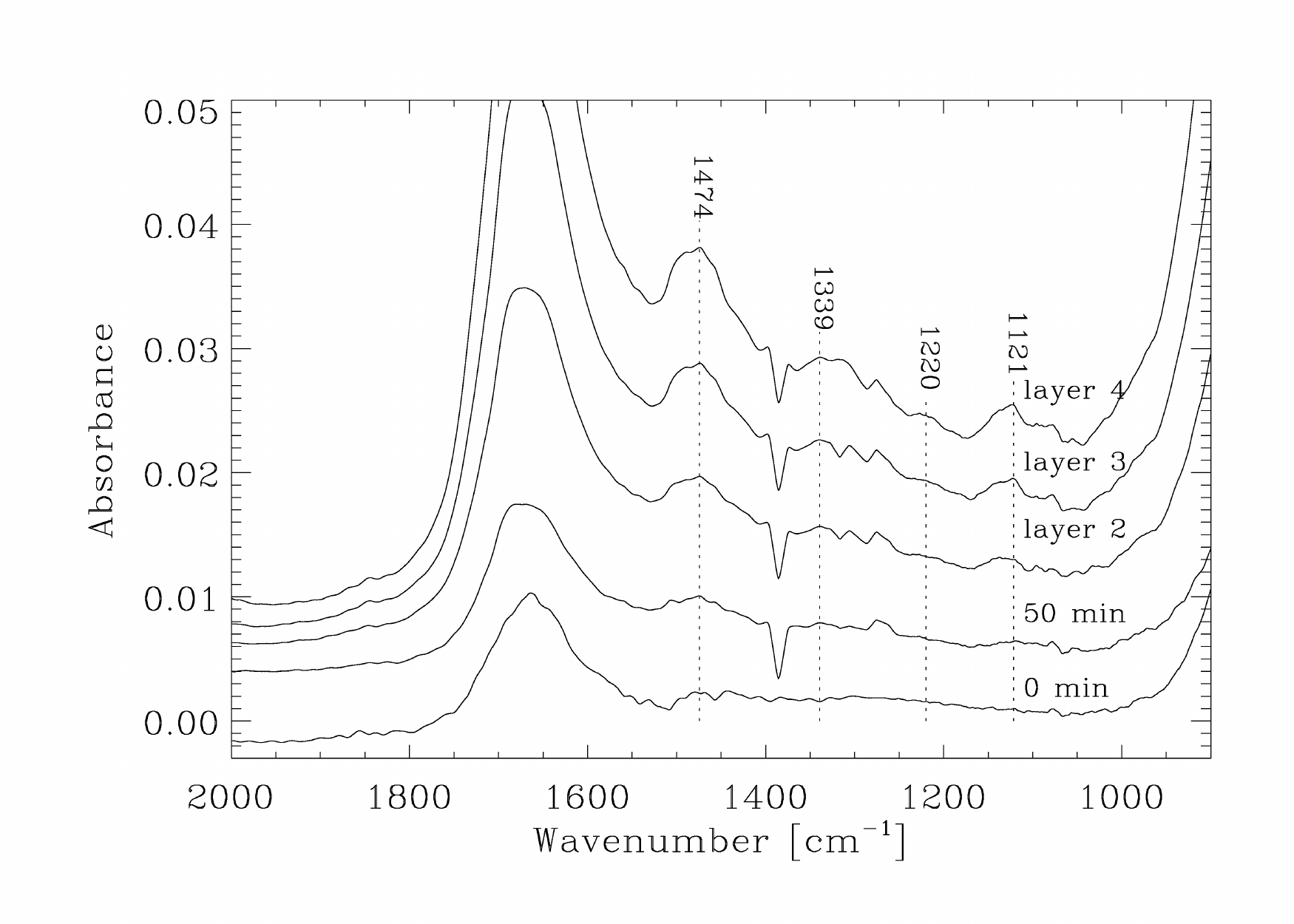} 
      \caption{Infrared spectra of H$_2$O:HCN (2:1) after 50 min of irradiation time. The spectra labelled as layer 2 to 4 correspond to new depositions on top of the previous one, each of them irradiated during 50 min.}
      \label{irrad}
\end{figure}

Infrared spectra of the H$_2$O:HCN (2:1) ice mixture, deposited at 8 $\mathrm{K}$, with an estimated column density of $N$ = 8 $\times$ 10$^{17}$ molecules cm$^{-2}$, and subsequently irradiated for 50 min, are shown in Fig.~\ref{irrad}. The average absorption cross section of water ice in the 120-160 nm range is 3.4 $\times$ 10$^{-18}$ cm$^2$ (Cruz-Diaz et al. 2014), a value compatible with Mason et al. 2006. This means that 95\% absorption of the incident UV flux corresponds to a column density of about 8 $\times$ 10$^{17}$ molecules cm$^{-2}$, and, therefore, each deposited ice layer was fully irradiated with an estimated dose of approximately 0.9 absorbed photon per molecule. 

Noble et al. (2013) proved that no thermal reaction is observed for a mixture of solid HCN:H$_2$O, neither at 10 K nor during heating to 180 K, where both HCN and H$_2$O desorbed.

During irradiation, HNCO, OCN$^-$, CO$_2$ and CO are formed as the main irradiation products, in agreement with the photoproducts reported by Gerakines et al. (2004).
Table~\ref{band_position_2} lists the products formed upon irradiation with its infrared band positions. From the top panel of Fig.~\ref{irrad} we can observe that the main feature of HNCO species is becoming clearly visible near 2262 cm$^{-1}$ as the irradiation time increases. This band is composed of three main components, at 2276, 2262, and 2245 cm$^{-1}$. As shown in Table~\ref{band_position_2}, the component higher in frequency can be assigned to the $^{13}$CO$_2$ species, while the remaining two components seem to be characteristic of the HNCO species (see also the HNCO spectra reported by Gerakines et al. 2004 and Khanna et al. 2002).

%%%%%%%%%%%%%%%%%%%%%%%%%%%%%%%%%%%%%%%%%%%%%%%%%%%%%%%%%
\begin{table*}
\centering
\caption{Infrared band position of photoproducts in an H$_2$O:HCN (2:1) ice mixture.}
\label{band_position_2}
\begin{tabular}{cc}   \hline \hline
\multicolumn{1}{c}{Wavenumber (cm$^{-1}$)} & Molecule \\ \hline
2342 & CO$_2$ \\
2276 & $^{13}$CO$_2$\\
2262 & HNCO (in H$_2$O environment)\\
2245 & HNCO (in HNCO environment)/HOCN$^a$\\
2170 & OCN$^-$\\
2138 & CO\\
2091 & HCN\\
1687 & HCONH$_2$$^b$\\
1612 & HCN (in H$_2$O:HCN)$^b$\\
1472 & NH$_4^+$$^b$\\
1314 & HCONH$_2$$^b$\\
1220 & HOCN$^a$\\
1121 & NH$_3$$^b$\\
\hline
\end{tabular}
\begin{list}{}
\item $^a$ Tentative assignment. 
\item $^b$ From van Broekhuizen et al. (2004).
\end{list}
\end{table*}
%%%%%%%%%%%%%%%%%%%%%%%%%%%%%%%%%%%%%%%%%%%%%%%%%

\begin{figure} [ht!]
   \centering
    \includegraphics[width=7.5cm]{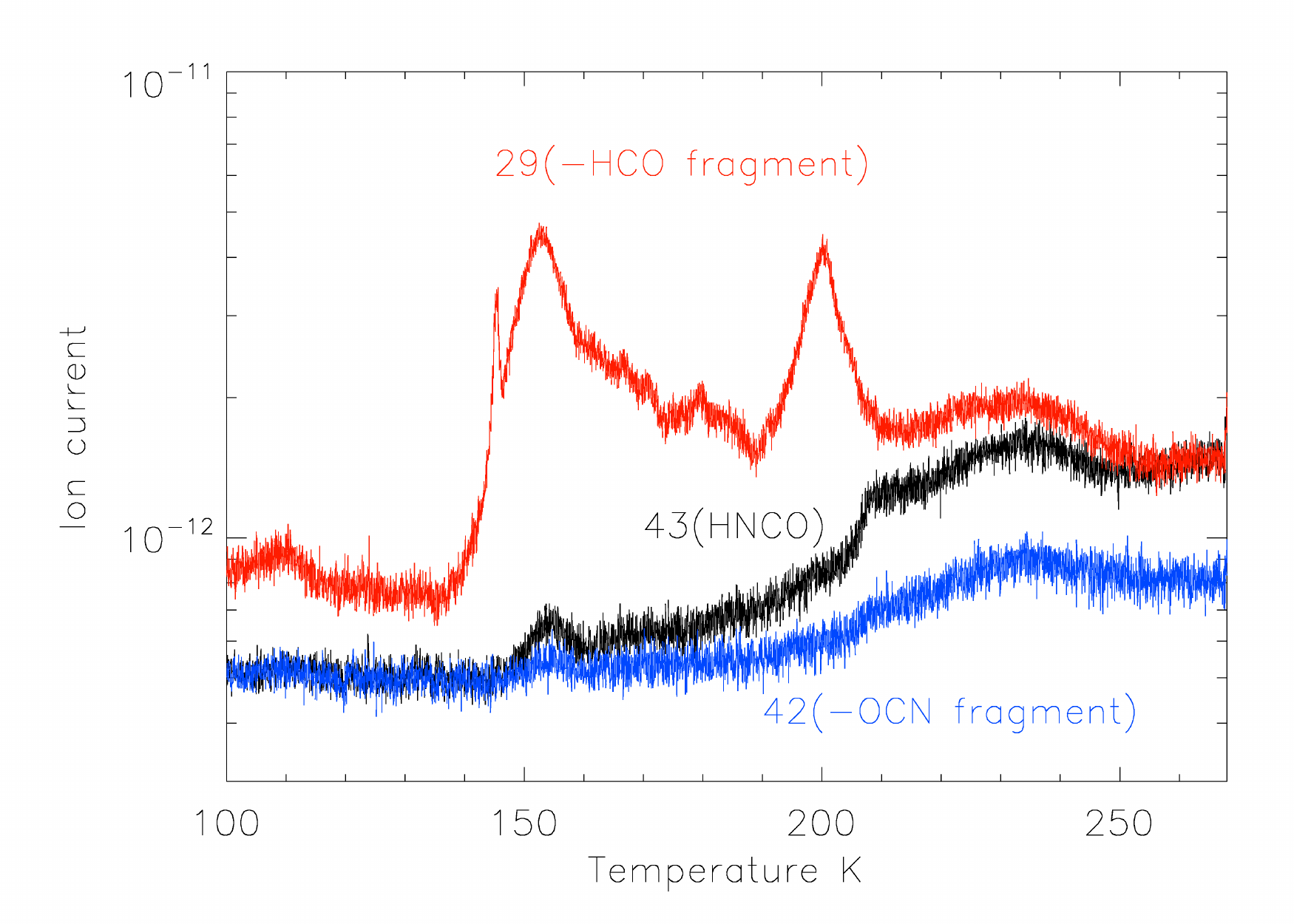}
    \includegraphics[width=7.5cm]{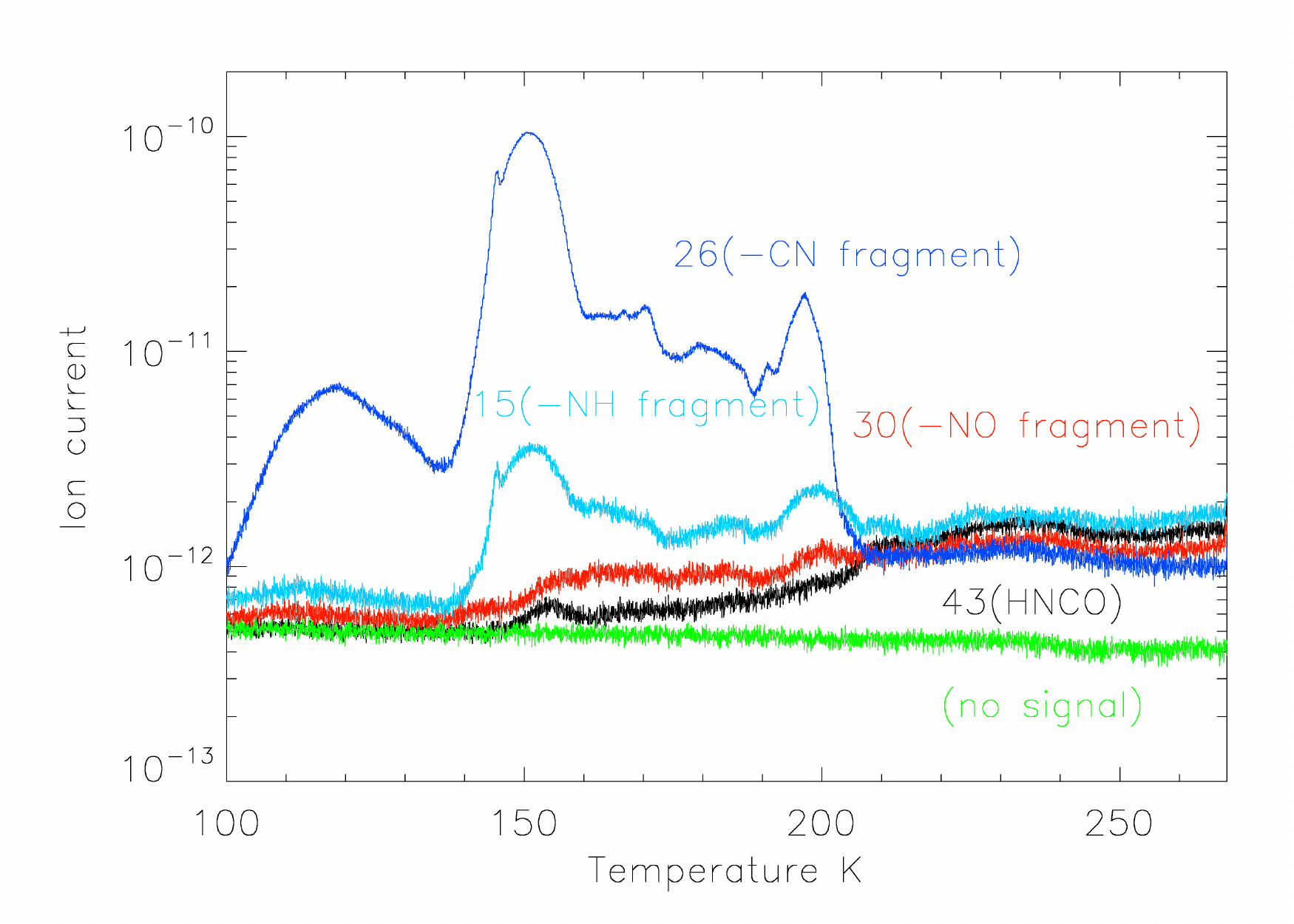}     
      \caption{Thermal desorption of UV-irradiated H$_2$O:HCN (2:1). The m/z~=~43 value refers to the molecular ion of the HNCO species and its isomers. The figure is split in two panels: the top panel shows the masses corresponding to the most abundant fragments of the HNCO isomer; the bottom panel shows the potential fragments of other isomers that might be present in the ice.}
      \label{qms_HCN}
\end{figure}

During warm-up of the irradiated H$_2$O:HCN ice mixture, the TPD curves of the m/z values for the HNCO molecular ion (m/z~=~43) and its fragments NCO$^+$ (m/z~=~42), and HCO$^+$ (m/z~=~29) have been recorded. The corresponding data are shown in Fig.~\ref{qms_HCN}. The comparison with the TPD data recorded from the experiments with the H$_2$O:CO:NH$_3$ ice mixture will be discussed in the next section. 

The products detected in the gas phase during the warm-up were not observed as photodesorbed products during the irradiation of the ice prior to warm-up.

%%%%%%%%%%%%%%%%%%%%%%%%%%%%%%%%%%%%%%%
\section{Discussion}
\label{disc}

\indent\indent The following discussion will focus mainly on the production of the HNCO species by irradiation of the ice mixtures and the possible formation of higher energy isomers. These processes can occur during the photoprocessing or can be induced by subsequent thermal processes or thermal desorption.

In the H$_2$O:NH$_3$:CO ice mixture, the low signal of the HNCO species in both infrared and mass spectra, complicates the data analysis and it is difficult to estimate the contribution of the different isomers to the curve profiles. For this reason, this analysis will be made on the basis of the data of the H$_2$O:HCN ice mixture.

In this mixture, the main feature around 2260 cm$^{-1}$ of the HNCO species became clearly visible in the infrared spectra as the irradiation time and deposited layers increased (see top panel of Fig.~\ref{irrad}). The two  characteristic components of this band (2262 and 2245 cm$^{-1}$) can be assigned to different environments, corresponding to a HNCO molecule in a H$_2$O or in a HNCO environment (see Theule et al. 2011 and references therein). Alternatively, the different components can be seen as an indication of the presence of crystallization processes in the ice.  From this feature alone it is difficult to discriminate between these two possibilities.

Teles et al. (1989) report the presence of a band characteristic of the HOCN isomer in the infrared spectrum of the matrix isolated species at 1228 cm$^{-1}$, which is the second most intense band with a relative intensity of 0.92 with respect to the main band. The bottom panel of Fig.~\ref{irrad} shows the 2000-900 cm$^{-1}$ spectroscopic region for the H$_2$O:HCN irradiated ice. In the fourth layer, after 50 min of irradiation time, a small feature at \emph{ca.} 1220 cm$^{-1}$ become visible. This apparent absorption can be regarded as a tentative identification of the HOCN species formed in the photoprocessed ice.

We estimated the abundance of the HNCO isomer and a rough estimation of the HOCN abundance formed in the ice on the basis of their column densities calculated as indicated in Sect.~\ref{expe}. The bands that have been considered for this analysis are the transitions at 2262 and 1220 cm$^{-1}$, for the HNCO and HOCN isomers, respectively, which have been assigned to the CN stretching (2262 cm$^{-1}$) and COH bending (1220 cm$^{-1}$). Using the band strength value of 7.2 $\times$ 10$^{-17}$ cm molecule$^{-1}$ from van Broekhuizen et al. (2004) for the ${\nu}_{\rm CN}$(HNCO) vibration, a column density of 2.2 $\times$ 10$^{16}$ cm$^{-2}$ has been derived for the HNCO species. No information about the band strength for the ${\delta}_{\rm COH}$(HOCN) transition is available. We attempted an estimation of its value based on the relative intensities of the two transitions from Teles et al. (1989), as indicated above, and using as an approximation that the band strength of the CN stretching is similar for both isomers, following the relation\\
\begin{equation}
A_{{\delta}_{\rm COH}{\rm (HOCN)}}=0.92\times{A_{{\nu}_{\rm CN}{\rm (HOCN)}}}=0.92\times{A_{{\nu}_{\rm CN}{\rm (HNCO)}}}
\label{relint}
\end{equation} 
from which a value of 6.6 $\times$ 10$^{-17}$ cm molecule$^{-1}$ has been derived. Using this value, an upper limit of the relative abundance of at most 2\% of HOCN with respect to the most stable isomer was estimated.

The tentative assignment of the second most stable isomer, HOCN, is also supported by the results of the mass spectra from the TPD experiments. As mentioned above, in the mass spectra of organic compounds the examination of the fragmentation pattern of the molecule, together with the information from the molecular ion, can be indicative of its structure. The two isomers, HNCO and HOCN, are supposed to have different fragmentation patterns (see Fischer et al. 2002, and Hop et al. 1989). In Fig.~\ref{qms_HCN} are shown the results of the thermal desorption of the H$_2$O:HCN irradiated ice at increasing temperature. 

It is interesting to notice the similarity of the desorption profiles for the species with m/z~=~15, 26, 30, and 43 (molecular ion of HNCO and its isomers) recorded at T~$>$~200 $\mathrm{K}$, where the OCN$^-$NH$_4^+$ species starts to desorb as NH$_3$ and HNCO isomers, and water is already desorbed.

\begin{figure} [ht!]
   \centering
    \includegraphics[width=7.5cm]{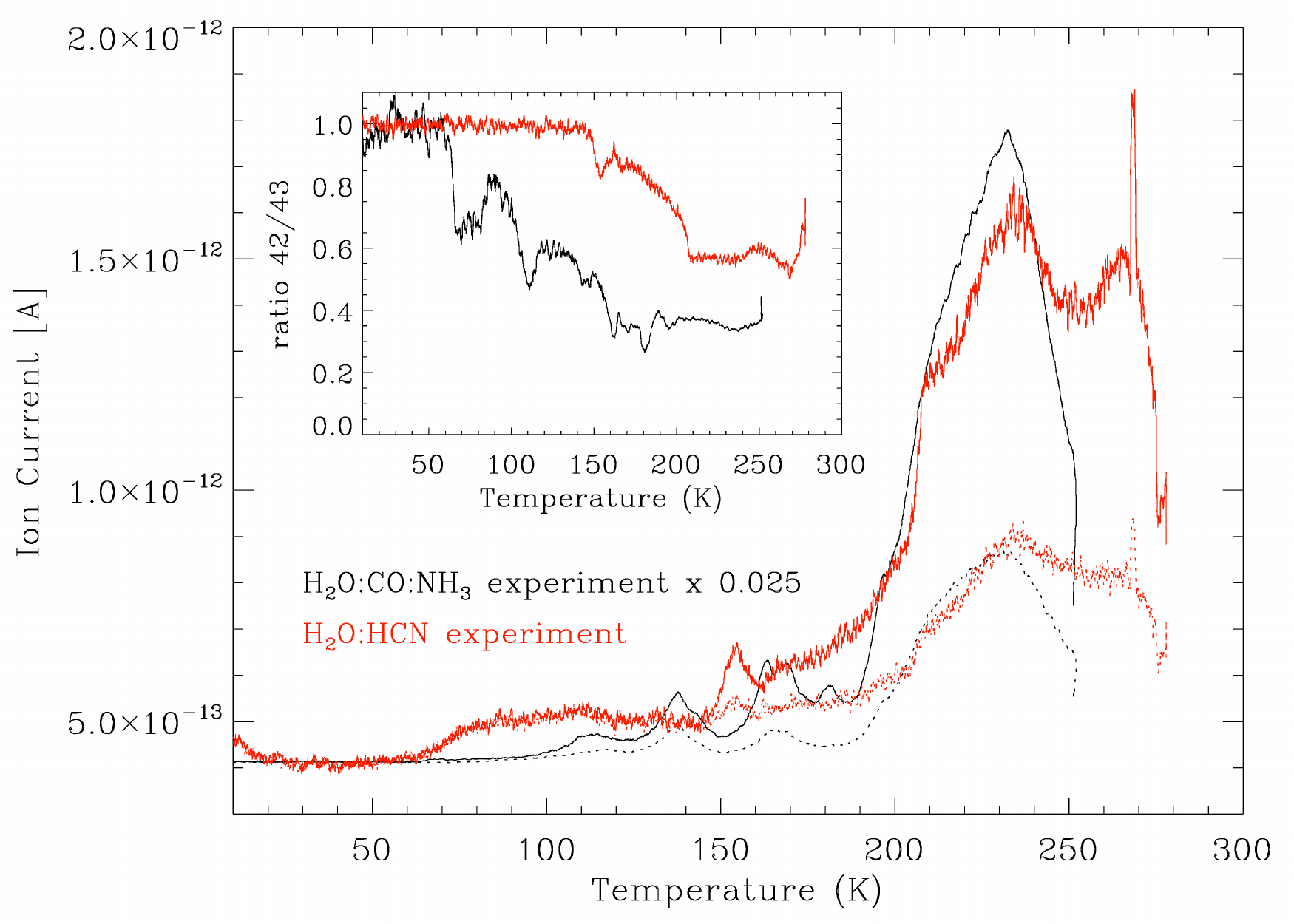}   
      \caption{Thermal desorption of UV-irradiated H$_2$O:NH$_3$:CO (black) and H$_2$O:HCN (red) ice mixtures. Solid lines correspond to m/z = 43, dotted lines correspond to m/z = 42.}
      \label{qms}
\end{figure}

Furthermore, the analysis of the fragmentation ratio of the HNCO species has been used to interpret the results of the ice mixture desorption. The TPD curves profiles for the H$_2$O:NH$_3$:CO (2:1:1) and H$_2$O:HCN (2:1) experiments are shown in Fig.~\ref{qms}. During the heating, in the H$_2$O:HCN experiment, the 42/43 mass ratio changed to 0.6 at temperatures above 220 $\mathrm{K}$; while in the H$_2$O:NH$_3$:CO experiment the mass ratio changed to approximately 0.3. The typical mass ratio of 0.2, which corresponds to the HNCO isomer, has never been observed during the ice desorption. This is an indication of the presence of other species contributing to the observed ratio. Since no other known products in the experiments contribute to the m/z = 42, 43, the most likely explanation is that the isomer desorbing is HOCN, which is also formed by the OCN$^-$ protonation. At the present stage, the fragmentation pattern of the HNCO species, using QMS, is unknown. Therefore, a quantitative estimation of the isomer abundances was not possible.

The search for the higher energy isomers HCNO and HONC in the ice mixture following VUV irradiation is not trivial. The matrix isolated infrared spectra of HCNO in various matrices, reported by Teles et al. (1989), show that its most intense absorption (the ${\nu}_{\rm CN}$) is around 2200 cm$^{-1}$, varying the position depending on the matrix. In our experiments this band overlaps with the very strong absorption of OCN$^{-}$ and cannot be observed. The COH bending (${\delta}_{\rm COH}$, around 1240 cm$^{-1}$) has a relative intensity too low to be detected in the ice mixture in a small amount.

On the contrary, the least stable isomer, HONC, could be observable in the ice from its infrared spectrum, because the ${\delta}_{\rm COH}$ measured at 1232 cm$^{-1}$ in an argon matrix has a relative intensity of \emph{ca.} one third of the most intense band, comparable with the one observed for HOCN. Unfortunately, the spectral position of the ${\delta}_{\rm COH}$ for the two species is very close and, therefore, the discrimination between them was not possible. Furthermore, the fragmentation pattern for HONC will probably produce the same fragments, m/z = 17 ($-$OH) and m/z = 26 ($-$CN), as for HOCN. Hence, from the infrared and mass spectrometric data, it is not possible to differentiate between the HONC and the HOCN species. Using the experimental techniques presented in this paper, it is, therefore, not possible to know if HCNO and HONC isomers are produced in our experiments.

The role of the OCN$^-$ species as a reaction intermediate in the formation of the HNCO isomers is not clear in our experiments. The band at 2170 cm$^{-1}$, characteristic of OCN$^-$, is increasing during the irradiation simultaneously with the spectroscopic features characteristic of the HNCO isomers, suggesting that the OCN$^-$ can be seen as an irradiation product of the ice rather than as an intermediate species. The small amount of HNCO isomers produced in the ice, however, makes the investigation of the kinetic processes extremely difficult.

%%%%%%%%%%%%%%%%%%%%%%%%%
\section{Astrophysical implications}
\label{astro}

\indent\indent The presence of  HCN in interstellar/circumstellar ice has not been confirmed, but this molecule is observed widespread in the gas phase and in comets. The HNCO:HCN observed ratio in the gas-phase is estimated to be $\sim$ 2\%, which is comparable with the estimation obtained in our experiments. Indeed, for the most favorable ratio of H$_2$O:HCN (2:1) we obtained a HNCO:HCN ratio of 3\%. 

HNCO and HOCN have been observed both in cold and warm molecular clouds. The ratio HOCN:HNCO $\leq$ 2\% in our experiment with irradiated H$_2$O:HCN ice is compatible with the 1\% value reported by Br\"{u}nken et al. (2010) toward Sgr B2. These species did not photodesorb during the ice irradiation. This could indicate that the ejection of the HNCO isomers from ice mantles in cold environments, like dark clouds, requires a different non-thermal desorption process or gas-phase formation mechanisms only.

Alternatively, gas-phase processes could be the main channels leading to formation of HNCO and its isomers in cold clouds. In particular, since HCNO has only been detected in cold dark cores and not observed in warm clouds, it was suggested to be formed in the gas-phase (Marcelino et al. 2009, 2010). Gas-grain chemical models reproduce reasonably well the observed abundances of HCNO and HOCN in both cold and warm clouds, which are formed through a combination of gas-phase and grain processes (Marcelino et al. 2010; Quan et al. 2010). The highest energy HNCO isomer, HONC, is predicted to be produced only on grain mantles in Quan et al. (2010) models. Due to the limitations of the experimental techniques, it was not possible to search for this isomer in our experiments. Its detection in space, whether in cold or warm environments, will provide further information to constrain the production of HNCO isomers.

%%%%%%%%%%%%%%%%%%%%%%%%%%%%%%%%%%%%%%%%%%%%%
\section{Conclusion}
\label{conc}

\indent\indent In this article we presented a study about the photoprocessing of interstellar ice analogues to investigate the formation of the OCN$^-$ ion, the HNCO species and its high-energy isomers. 
Several experiments of ice irradiation followed by warm-up were conducted, including the more abundant molecular components observed in ice mantles embedded in the H$_2$O-ice matrix, i.e. a carbon-bearing molecule (CO, CH$_3$OH, or CH$_4$) and a nitrogen source (NH$_3$ or N$_2$). Information about the different ice mixtures investigated in the present work are summarized in Table~\ref{summary}. Of these, only the H$_2$O:NH$_3$:CO ice mixture led to a detectable amount of HNCO. Inclusion of the HCN molecule in the H$_2$O-ice matrix allowed the formation of HNCO and an upper limit of 2\% relative to HNCO was derived for the HOCN isomer, as inferred from IR and QMS data.

%%%%%%%%%%%%%%%%%%%%%%%%%%%%%%%%%%%%%%%%%%%%%%
\begin{table} [tb]
\centering
\caption{Summary of the different mixtures investigated in the present work.$^a$}
\label{summary}
\begin{tabular}{ccc}   \hline \hline
Ice mixture & Ratio & HNCO formation\\ \hline
CH$_3$OH:NH$_3$ & 1:1 & No \\
H$_2$O:CH$_4$:N$_2$ & 1:1:1 & No \\
H$_2$O:CO:N$_2$ & 2:1:1 & No \\
NH$_3$:CO & 1:1 & No \\
H$_2$O:NH$_3$:CO & 2:1:1 & Yes$^b$ \\
H$_2$O:HCN & 1:1 & No \\
H$_2$O:HCN & 2:1 & Yes \\
H$_2$O:HCN & 3:1 & No \\
\hline
\end{tabular}
\begin{list}{}
\item $^a$ Ratios estimated from the recorded spectra.
\item $^b$ Its formation can only be inferred from the QMS data.
\end{list}
\end{table}
%%%%%%%%%%%%%%%%%%%%%%%%%%%%%%%%%%%

In both ice mixtures the OCN$^-$ species is the most abundant photoproduct. However, from the present experiments, it is not possible to investigate its role in the formation and isomerization of HNCO species.
The ultra-high vacuum conditions of the ISAC system allowed the detection of very small amounts of HNCO isomer, in line with our previous works on other species (e.g., Jim\'enez-Escobar et al. 2012).
Below, we discuss that, despite the low formation efficiency of HNCO isomers in our experiments, it may be sufficient to account for the abundances detected in space of HNCO and HOCN.

Upper limits of the expected abundances of these species relative to hydrogen in regions where icy grain mantles are submitted to irradiation can be roughly estimated assuming that the abundance of H$_2$O is of the order of 10$^{-4}$, for HCN that was not detected in the ice we can take 1 $\times$ 10$^{-6}$ as an upper limit, then the corresponding HNCO and HOCN maximum abundances are about 3 $\times$ 10$^{-8}$ and 6 $\times$ 10$^{-10}$, see Sect.~\ref{astro}, which are compatible with the observed abundances in the gas phase.

%%%%%%%%%%%%%%%%%%%%%%%%%%%%%%%%%%%%%%%%%%%%%%%%%%%%%%%%%%
\section*{Acknowledgements}

We thank C. Menor Salv\'an for his advice on the synthesis of HCN. A. J. was supported by a training grant from INTA. B.~M.~G. acknowledges M. Castellanos for his assistance and support from CONSOLIDER grant CSD2009-00038. This research was financed by the Spanish MICINN under Project AYA2011-29375 and CONSOLIDER grant CSD2009-00038. N. M. acknowledges The National Radio Astronomy Observatory(NRAO). NRAO  is a facility of the National Science Foundation operated under cooperative agreement by Associated Universities, Inc.

%%%%%%%%%%%%%%%%%%%%%%%%%%%%%%%%%%%%%%%%%%%%%%%%%%%%%%%%%%

\end{document}